\documentclass[12pt]{article}
\usepackage{epsfig,picinpar,floatflt,amssymb}
\newcommand{\resection}[1]{\setcounter{equation}{0}\section{#1}}
\newcommand{\appsection}{\addtocounter{section}{1} \setcounter{equation}{0}
             \section*{Appendix \Alph{section}}}

\def\to{\rightarrow}
\def\goto{\longrightarrow}

\def\al{\alpha}
\def\th{\theta}

\def\bd{\begin{displaystyle}}
\def\ed{\end{displaystyle}}
\def\ba{\begin{array}}

\def\ea{\end{array}}
\def\EQ{\begin{equation}}
\def\EN{\end{equation}}
\def\bea{\begin{eqnarray}}
\def\eea{\end{eqnarray}}
\def\beano{\begin{eqnarray*}}
\def\eeano{\end{eqnarray*}}

\def\hs{\hspace{0.1in}}

\begin{document}
\oddsidemargin 5mm
\setcounter{page}{0}
\newpage     
\setcounter{page}{0}
\begin{titlepage}
\begin{flushright}
\end{flushright}
\vspace{0.5cm}
\begin{center}
{\large {\bf Universal amplitude ratios \\
in the two-dimensional $q$-state Potts model and percolation \\
from quantum field theory}}\footnote{Work supported 
in part by the European Union under contract FMRX-CT96-0012} \\
\vspace{1.5cm}
{\bf G. Delfino$^{a}$ and J.L. Cardy$^{b,c}$} \\
\vspace{0.8cm}
$^a${\em Laboratoire de Physique Th\'eorique,
Universit\'e de Montpellier II \\
Pl. E. Bataillon, 34095 Montpellier, France} \\
$^b${\em Theoretical Physics, University of Oxford\\
1 Keble Road, Oxford OX1 3NP, United Kingdom} \\ 
$^c${\em All Souls College, Oxford} \\
\end{center}
\vspace{6mm}
\begin{abstract}
\noindent
We consider the scaling limit of the two-dimensional $q$-state Potts model for
$q\leq 4$. We use the exact scattering theory proposed by Chim and 
Zamolodchikov to determine the one and two-kink form factors of the energy,
order and disorder operators in the model. Correlation functions and universal
combinations of critical amplitudes are then computed within the two-kink
approximation in the form factor approach. Very good agreement is found 
whenever comparison with exact results is possible. We finally consider the 
limit $q\to 1$ which is related to the isotropic percolation problem. 
Although this case presents a serious technical difficulty, we predict a value 
close to 74 for the ratio of the mean cluster size amplitudes above and 
below the percolation threshold. 
Previous estimates for this quantity range from 14 to 220.
\end{abstract}
\vspace{5mm}
\end{titlepage}
\newpage

\setcounter{footnote}{0}
\renewcommand{\thefootnote}{\arabic{footnote}}

\newpage
\resection{Introduction}
Two-dimensional integrable models enjoy a very particular role in the framework
of quantum field theory. As a matter of fact, they provide the only 
known examples of non-trivial, interacting relativistic quantum theories to be
exactly solved. As is well known, this circumstance has to be traced back to 
the existence in these models of non-trivial integrals of motions resulting 
in a strong simplification of the scattering theory and in the possibility of
determining the exact $S$-matrix through bootstrap techniques. A large number 
of
integrable models has been discovered and solved in this way over the last 
two decades \cite{ZZ,Sasha,Giuseppe}.

If the theoretical relevance of integrable field theories is obvious, surely 
their effectiveness for the accurate quantitative description of interesting
physical systems is not less important. In this respect, two-dimensional 
statistical mechanics offers a natural testing 
ground. Many statistical mechanical
models are known to be exactly solvable in two dimensions \cite{Baxter} and 
are then natural candidates for a description in terms of integrable field 
theory in the scaling limit in which the correlation length becomes much 
larger than the lattice spacing. Moreover, one of the remarkable 
results of the recent investigations of two-dimensional quantum 
field theory is that integrability can be a property of the scaling limit 
of statistical models even in cases where it has never been found on the 
lattice. The critical Ising model in a magnetic field provides the most known
example of such a  situation\footnote{It is interesting to notice that for 
this specific case a lattice model has been identified which is in the same
universality class than the Ising model in a magnetic field and is integrable
\cite{WNS}.}\cite{Sasha}. 

It is clear that, for the purpose of the quantitative study of statistical 
systems through the methods of quantum field theory, only universal quantities
should be considered, i.e. those quantities which do not depend on the 
particular microscopic realisation of the system and are determined instead 
by its global features (essentially, internal symmetries and dimensionality).
Universal properties emerge when the system approaches a second order phase
transition point. For a magnetic system in zero external field, the usual
characterisation of criticality is in terms of the classical 
thermodynamic quantities: in the vicinity of the critical point the specific 
heat, spontaneous magnetisation, susceptibility and correlation length behave 
as
\bea
&& C\simeq (A_\pm/\alpha)\,\tau^{-\alpha}\,, \nonumber\\
&& M\simeq B\,(-\tau)^{\beta}\,, \nonumber\\
&& \chi\simeq\Gamma_\pm\,\tau^{-\gamma}\,, \nonumber\\
&& \xi\simeq\xi^\pm_0\,\tau^{-\nu}\,,
\label{amplitudes}
\eea
where we denoted by $\tau$ the reduced temperature, $\tau=a(T-T_c)$, $a>0$,
and the labels $\pm$ on the critical amplitudes refer to the critical point
being approached from above or from below. While the critical exponents are 
characteristic of the critical point and in $d=2$ are determined by conformal 
field theory (CFT) \cite{BPZ,ISZ}, the critical amplitudes carry information
about the renormalisation group trajectory along which the critical point is 
approached and their determination requires a study of the system away from 
criticality. Critical amplitudes depend on metric factors and are not 
themselves universal, but universal combinations of them can by constructed 
which characterise the scaling region around the critical point. The following 
universal amplitude ratios are usually considered in the literature \cite{PHA}
\bea
&& A_+/A_-\,,\hspace{.5cm}\Gamma_+/\Gamma_-\,,\hspace{.5cm}\xi_0^+/\xi_0^-\,, 
\nonumber\\
&& R_C=A_+\Gamma_+/B^2\,,\hspace{.5cm}R_\xi^+=A_+^{1/d}\xi^+_0\,\,.
\label{ratios}
\eea
The scale factor independence of $R_C$ and $R_\xi^+$ is a direct consequence
of the scaling and hyperscaling relations
\EQ
\al=2-2\beta-\gamma\,,\hspace{1cm}2-\al=d\nu\,\,.
\label{scaling}
\EN

The critical amplitudes can be expressed as moments of correlation functions 
of the spin and energy operators in the off-critical theory. As a consequence,
the computation of the amplitude ratios in the framework of integrable field 
theory requires bridging the gap between the scattering theory, in terms of 
which the solution of the model is given, and the off-shell dynamics. With
respect to this problem, the so called form factor bootstrap is the method 
which proved so far as the most effective. In this approach, correlation 
functions are expressed as spectral sums over $n$-particle intermediate states.
The operator matrix elements entering the decomposition (known as form factors)
are subject to a set of monodromy and residue equations \cite{KW,Smirnov} which
have been solved exactly for many integrable models. These equations, however,
are fixed by the $S$-matrix alone and do not distinguish between different 
operators. Although the selection rules coming from internal symmetries
together with some minimality assumptions are sufficient in some cases to
identify the form factor solutions corresponding to specific operators, it 
turns out that in general more information about the operator space has to
be injected in order to handle this problem \cite{immf,DS}. 

The operators which appear in physical applications are the scaling operators. 
Since they are naturally defined in the scaling limit towards the ultraviolet
fixed point, it is not surprising that the constraints for their identification
in the form factor approach come from the high energy asymptotics of the 
matrix elements. Two such constraints have been identified which are crucial
for the results obtained in this paper. The first one relates the asymptotic
behaviour of form factors of the scaling operators to their scaling dimension
\cite{immf}; the second constraint takes the form of a cluster decomposition of
the matrix elements when the momenta of a subset of particles become much
larger than the others. It was argued in Ref.\,\cite{DSC} that the latter 
property is related to the decoupling of the theory into holomorphic and 
anti-holomorphic sectors in the ultraviolet limit. 

This level of understanding proved so far sufficient for the determination of 
the two-particle form factors of physically relevant operators in integrable 
theories. From the general point of view, the computation of the two-particle 
matrix elements amounts to fixing the initial conditions of the bootstrap 
procedure. The determination of the form factors with more than two particles 
through the residue equations
is a mathematical problem which can be straightforward or extremely difficult 
depending on the degree of complexity of the underlying scattering theory. 
Although the problem has been solved in several specific cases (see e.g. 
\cite{BKW,Smirnov,ZamYL}), no general method of solution is known for the case 
when the scattering allows for the exchange of quantum numbers among particles
with the same mass. 

If in principle this circumstance severely restricts the range of applicability
of the form factor approach for the exact computation of correlation 
functions,
in practice very accurate results can be obtained anyway in a large number 
of cases. In fact, it has been by now  verified for several integrable models
that the spectral series over form factors is characterised by a remarkably 
fast rate of convergence (see, among others, Refs.\,\cite{YZ,ZamYL,CM2,immf}).
A theoretical justification of this property was proposed in \cite{CM2}
relying on phase space considerations and some peculiarities of integrable 
dynamics. As a matter of fact, it turns out that zeroth moments of two-point
correlators can be computed within a typical accuracy of order 1\%, including
in the spectral series the one and two-particle contributions only. At fixed
level of approximation, the accuracy rapidly increases if higher moments 
are considered in which the contribution coming from the short distances (more 
sensitive to the exclusion from the sum of many-particle states) is suppressed.

It is the purpose of this paper to use the form factor approach to compute the 
universal amplitude ratios for the $q$-state Potts model ($q\leq 4$) and 
the isotropic percolation problem. Although the latter does not fit immediately
in the standard terminology of thermal phase transitions, it can be formally
related to the $q\to 1$ limit of the Potts model. Through this mapping, the 
thermodynamic quantities listed in (\ref{amplitudes}) become related to the 
mean number of clusters, the percolation probability, the mean cluster size
and the pair connectivity, respectively. 

Our starting point will be the exact scattering theory for the $q$-state Potts
model proposed by Chim and Zamolodchikov in Ref.\,\cite{CZ}. Due to the fact 
that the fundamental constraint of permutation symmetry which characterises the
model is very naturally imposed on the interaction of the kinks interpolating 
between the $q$ degenerate vacua of the spontaneously broken phase, the 
scattering theory is formulated at $T<T_c$. 
Form factors over kinks have not been previously considered in the literature.
Although most of the formalism used to deal with ordinary particles goes 
through with minor adaptations, some interesting new features appear. This 
happens, in particular, when operators which are non-local with respect to the 
kinks (e.g. the magnetisation operator) are considered. It turns out that the 
way non-locality is implemented in the low-temperature formalism is markedly 
different from that characteristic of the unbroken phase (see \cite{YZ} for a 
review of the latter).

The amplitude ratios (\ref{ratios}) involve both high and low-temperature 
amplitudes. We will work all the time in the low-temperature phase and rely on
duality to get the information about the unbroken phase. In particular, the 
correlators of the magnetisation at $T>T_c$ will be obtained computing those 
of the disorder operators at $T<T_c$. An interesting problem, however, arises 
at this point. As the form factors of the magnetisation and disorder 
operators are determined quite independently from each other, the 
relative normalisation between the two operators remains unfixed. This 
needs to be determined if the two operators are to describe the
same physical
quantity in the two different phases. The obvious solution to this problem
would be to compute the two-point correlators of the two operators and to match
the coefficients of their short distance asymptotics. In practice this cannot
be done even in the cases in which all the multi-particle form factors are 
known, simply because nobody knows how to resum the spectral series. Once 
again the solution is provided by the asymptotic properties of form factors
and in particular by the cluster factorisation mentioned above: a two-kink
magnetisation matrix element has to factorise into the product of two one-kink
disorder form factors, and this requirement fixes the relative normalisation.

We will compute correlation functions and their moments in the $q$-state Potts
model within the two-kink approximation in the form factor approach.
When comparison with exact results is possible, the fast convergence of the 
spectral series is confirmed and results with the typical accuracy mentioned
above are obtained. A serious technical difficulty however
arises in the computation
of the two-kink form factor of the magnetisation operator. Here, the conspiracy
of non-locality and non-diagonal scattering results in a new form of monodromy
problem expressed by a functional equation that we are not able to solve for
generic values of $q$. The problem can be overcome for $q=2,3,4$ and the 
complete list of amplitude ratios (\ref{ratios}) will be presented for this 
cases. For percolation, however, we can give accurate results only for some 
universal ratios. For the others we propose an extrapolation in $q$ which,
although naive, leads to results in good agreement with the available 
lattice estimates (series enumerations, Monte Carlo). Quite particular is 
the case of the ratio of the cluster size amplitudes $\Gamma_+/\Gamma_-$.
Here the existing estimates range from 14 to 220, and our extrapolated value
$\approx74$ represents a considerable improvement in accuracy.

The layout of the paper is the following. In the next section we briefly 
review the description of the scaling limit of the $q$-state Potts model as a 
perturbed CFT and the associated scattering theory. The one and two-particle 
form factors of the energy, order and disorder operators are determined in 
section 3. The results given by the form factor approach in the 
two-kink approximation for the central charge and scaling dimension sum rules, 
and for the universal amplitude ratios in the Potts model, are 
presented in section 4. We discuss percolation in section 5 before summarising 
our results and making some final remarks in section 6.

\resection{Scattering theory of the scaling Potts model}
The $q$-state Potts model \cite{Potts,Wu} is the generalisation of the Ising
model defined by the lattice Hamiltonian
\EQ
H=-J\sum_{(x,y)}\delta_{s(x),s(y)}\,,
\EN
where the sum is over nearest neighbours and the site variable $s(x)$ can 
assume $q$ possible values (colours). Clearly, the model is invariant under the 
group $S_q$ of permutations of the colours. In the ferromagnetic case $J>0$ we 
are interested in, the states in which all the sites have the same colour 
minimise the energy and the system exhibits spontaneous magnetisation at 
sufficiently low temperatures. There exists a critical temperature $T_c$ above 
which the thermal fluctuations become dominant and the system is in a 
disordered phase. If we introduce the variables
\EQ
\sigma_\al(x)=\delta_{s(x),\al}-\frac{1}{q}\,,\hspace{1cm}\al=1,2,\ldots,q
\label{sigma}
\EN
constrained by the condition
\EQ
\sum_{\al=1}^q\sigma_\al(x)=0\,,
\label{constraint}
\EN
the expectation values $\langle\sigma_\al\rangle$ differ from zero only in 
the low-temperature phase and can be used as order parameters.

After defining $x\equiv e^{J/T}-1$, the partition function of the model 
can be written in the form
\EQ
Z=\mbox{Tr}_s\prod_{(x,y)}(1+x\delta_{s(x),s(y)})\,\,.
\label{partition}
\EN
A graph ${\cal G}$ on the lattice can be associated to each Potts configuration
by drawing a bond between two sites with the same colour. In the above 
expression, a power of $x$ is associated to each bond in the graph. Taking 
into account the summation over colours one arrives to the expansion
\cite{Baxter2}
\EQ
Z=\sum_{\cal G}q^{N_c}x^{N_b}\,,
\label{newpartition}
\EN
where $N_b$ is the total number of bonds in the graph ${\cal G}$ and $N_c$
is the number of connected components (clusters) in ${\cal G}$ (each isolated
site is also counted as a cluster). In terms of the partition function
(\ref{newpartition}) the $q$-state Potts model is well defined even for 
noninteger values of $q$.

In two dimensions, the Potts model is known to undergo a first order phase 
transition at $T=T_c$ for $q>4$ \cite{Baxter}. For $q\leq 4$, 
however, the transition is continuous and the critical point can be described
by a CFT. Relying on the knowledge of some critical exponents \cite{Nienhuis}, 
Dotsenko and Fateev identified the central charge of this CFT to be \cite{DF}
\EQ
c=1-\frac{6}{t(t+1)}\,,
\label{c}
\EN
where the parameter $t$ is related to $q$ by the formula
\EQ
\sqrt{q}=2\sin\frac{\pi(t-1)}{2(t+1)}\,\,.
\EN
The scaling dimension $x_\sigma$ of the magnetisation operators $\sigma_\al(x)$
(the continuous version of the quantities (\ref{sigma})) is identified with
that of the primary operator $\phi_{(t-1)/2,(t+1)/2}$ in the CFT
\EQ
x_\sigma=\frac{(t-1)(t+3)}{8t(t+1)}\,\,.
\label{xsigma}
\EN
The energy density operator $\varepsilon(x)$ ($\sim\sum_y\delta_{s(x),s(y)}$ 
on the lattice) coincides with the primary operator $\phi_{2,1}$ with scaling
dimension
\EQ
x_\varepsilon=\frac{1}{2}\left(1+\frac{3}{t}\right)\,\,.
\EN
In view of these identifications, the continuum limit of the $q$-state Potts 
model ($q\leq 4$) is described by the perturbed CFT 
\EQ
{\cal A}={\cal A}_{CFT}+\tau\int d^2x\,\varepsilon(x)\,,
\label{action}
\EN
where ${\cal A}_{CFT}$ denotes the critical point action. Since a series of 
non-trivial integrals of motions is known to survive the deformation
of a CFT by the operator $\phi_{2,1}$ \cite{Sasha}, the off-critical theory
(\ref{action}) is integrable. This circumstance was exploited in 
Ref.\,\cite{CZ} to propose an exact scattering theory which we now briefly 
review\footnote{Notice that a different $S$-matrix for the $\phi_{2,1}$ 
perturbation of minimal models was determined by Smirnov \cite{Smirnov21}.
We use here the scattering description of Ref.\,\cite{CZ} because it is more
suitable for analytic continuation in $q$. Both descriptions must lead to the 
same results for the correlation functions. See \cite{Smirnovcomment} for a 
detailed discussion of this point in the case of the $\phi_{1,3}$ 
perturbation.}.

The low-temperature phase of the model is 
characterised by the presence of $q$ degenerate vacua that we label by the 
index $\alpha=1,2,\ldots,q$. The elementary excitations are then provided 
by kinks\footnote{We parameterise on-shell momenta as 
$p^\mu=(m\cosh\theta,m\sinh\theta)$, where $m$ denotes the mass of the kinks.} 
$K_{\alpha\beta}(\theta)$ interpolating between the two vacua $\alpha$ and 
$\beta$ ($\alpha\neq\beta$). The space of physical asymptotic states consists 
of multi-kink configurations of the type $K_{\alpha_0\alpha_1}(\th_1)
K_{\alpha_1\alpha_2}(\th_2)\ldots K_{\alpha_{n-1}\alpha_n}(\th_n)$ 
($\alpha_i\neq\alpha_{i+1}$) interpolating between the vacua
$\alpha_0$ and $\alpha_n$. As a consequence of the invariance under 
permutations, all the $n$-kink states fall into two topological sectors:
the ``neutral'' sector, corresponding to $\alpha_0=\alpha_n$, and the 
``charged'' sector, corresponding to $\alpha_0\neq\alpha_n$.

Integrability implies that the scattering processes are completely elastic and
factorised into the product of two-kink interactions. Since topological charge
is conserved, an outgoing two-kink state can only differ from the ingoing one 
by the vacuum state between the kinks.
Hence, the two-kink scattering can formally be 
described through the Faddeev-Zamolodchikov commutation relation
\EQ
K_{\al\gamma}(\th_1)K_{\gamma\beta}(\th_2)=\sum_{\delta\neq\al,\beta}
S_{\al\beta}^{\gamma\delta}
(\th_{12})K_{\al\delta}(\th_2)K_{\delta\beta}(\th_1)\,,
\label{fz}
\EN
where $\th_{12}\equiv\th_1-\th_2$ and $S_{\al\beta}^{\gamma\delta}(\th_{12})$ 
denotes the two-body scattering amplitude (Fig.\,1a). $S_q$-invariance reduce
to four the number of independent amplitudes, two for the charged and two
for the neutral topological sector
\bea
&& K_{\al\gamma}(\th_1)K_{\gamma\beta}(\th_2)=S_0(\th_{12})\sum_{\delta\neq
\gamma}K_{\al\delta}(\th_2)K_{\delta\beta}(\th_1)+S_1(\th_{12})
K_{\al\gamma}(\th_2)K_{\gamma\beta}(\th_1)\,\hspace{.5cm}\al\neq\beta
\nonumber\\
&& K_{\al\gamma}(\th_1)K_{\gamma\al}(\th_2)=S_2(\th_{12})\sum_{\delta\neq
\gamma}K_{\al\delta}(\th_2)K_{\delta\al}(\th_1)+S_3(\th_{12})
K_{\al\gamma}(\th_2)K_{\gamma\al}(\th_1)\,,
\eea
Using the commutation relation (\ref{fz}) twice one obtains the unitarity
constraint
\EQ
\sum_\varepsilon S_{\al\beta}^{\gamma\varepsilon}(\th)S_{\al\beta}
^{\varepsilon\delta}(-\th)=\delta^{\gamma\delta}\,,
\EN
which amounts to the set of equations
\bea
&& (q-3)S_0(\th)S_0(-\th)+S_1(\th)S_1(-\th)=1\,,\\
&& (q-4)S_0(\th)S_0(-\th)+S_0(\th)S_1(-\th)+S_1(\th)S_0(-\th)=0\,,\\
&& (q-2)S_2(\th)S_2(-\th)+S_3(\th)S_3(-\th)=1\,,\\
&& (q-3)S_2(\th)S_2(-\th)+S_3(\th)S_2(-\th)+S_2(\th)S_3(-\th)=0\,\,.
\eea
Crossing symmetry provides the relations 
\bea
&& S_0(\th)=S_0(i\pi-\th)\,\\
&& S_1(\th)=S_2(i\pi-\th)\,\\
&& S_3(\th)=S_3(i\pi-\th)\,\,.
\eea
Using these constraints together with the Yang-Baxter and bootstrap equations 
(that we do not need to reproduce here) the following expressions for the 
four elementary amplitudes were determined in Ref.\,\cite{CZ}
\bea
&& S_0(\th)=\frac{\sinh\lambda\th\,\sinh\lambda(\th-i\pi)}
{\sinh\lambda\left(\th-\frac{2\pi i}{3}\right)\,\sinh\lambda\left(\th-
\frac{i\pi}{3}\right)}\,\Pi\left(\frac{\lambda\th}{i\pi}\right)\,\\
&& S_1(\th)=\frac{\sin\frac{2\pi\lambda}{3}\,\sinh\lambda(\th-i\pi)}
{\sin\frac{\pi\lambda}{3}\,\sinh\lambda\left(\th-\frac{2 i\pi}{3}\right)}\,
\Pi\left(\frac{\lambda\th}{i\pi}\right)\,\\
&& S_2(\th)=\frac{\sin\frac{2\pi\lambda}{3}\,\sinh\lambda\th}
{\sin\frac{\pi\lambda}{3}\,\sinh\lambda\left(\th-\frac{i\pi}{3}\right)}\,
\Pi\left(\frac{\lambda\th}{i\pi}\right)\,\\
&& S_3(\th)=\frac{\sin\lambda\pi}{\sin\frac{\pi\lambda}{3}}\,
\Pi\left(\frac{\lambda\th}{i\pi}\right)\,,
\eea
where $\lambda$ is related to $q$ as
\EQ
\sqrt{q}=2\sin\frac{\pi\lambda}{3}\,,
\EN
and
\bea
&& \Pi(x)=-\frac{\Gamma(1-x)\Gamma(1-\lambda+x)\Gamma\left(\frac{7}{3}\lambda-
x\right)\Gamma\left(\frac{4}{3}\lambda+x\right)}
{\Gamma(1+x)\Gamma(1+\lambda-x)\Gamma\left(\frac{1}{3}\lambda+x
\right)\Gamma\left(\frac{4}{3}\lambda-x\right)}\,
\prod_{k=1}^{\infty}\Pi_k(x)\Pi_k(\lambda-x)\,,\nonumber\\
&& \Pi_k(x)=\frac{\Gamma(1+2k\lambda-x)\Gamma(2k\lambda-x)
\Gamma\left[1+\left(2k-\frac{1}{3}\right)\lambda-x\right]
\Gamma\left[\left(2k+\frac{7}{3}\right)\lambda-x\right]}
{\Gamma[1+(2k+1)\lambda-x]\Gamma[(2k+1)\lambda-x]
\Gamma\left[1+\left(2k-\frac{4}{3}\right)\lambda-x\right]
\Gamma\left[\left(2k+\frac{4}{3}\right)\lambda-x\right]}\,\,.\nonumber
\eea
We quote here the following integral representation of the function
$\Pi(x)$ which will be useful in the following
\bea
&&\Pi\left(\frac{\lambda\th}{i\pi}\right)=
\frac{\sinh\lambda\left(\th+i\frac{\pi}{3}\right)}{\sinh\lambda(\th-i\pi)}\,
e^{{\cal A}(\th)}\,,\\
&& {\cal A}(\th)=\int_0^\infty\frac{dx}{x}\,
\frac{\sinh\frac{x}{2}\left(1-\frac{1}{\lambda}\right)-
\sinh\frac{x}{2}\left(\frac{1}{\lambda}-\frac{5}{3}\right)}
{\sinh\frac{x}{2\lambda}\cosh\frac{x}{2}}\,\sinh\frac{x\th}{i\pi}\,\,.
\eea
It is easily seen that the function $\Pi(\lambda\th/i\pi)$ is free of poles 
in the physical strip $\mbox{Im}\th\in(0,\pi)$ for $q<3$ (i.e. $\lambda<1$).
Hence, in this range of $q$ the only poles of the scattering amplitudes in 
the physical strip are those located at $\th=2i\pi/3$ and $\th=i\pi/3$ and 
correspond to the appearance of the elementary kink itself as a bound state
in the direct and crossed channel, respectively. The residues
\bea
&& \mbox{Res}_{\th=2i\pi/3}S_0(\th)=-\mbox{Res}_{\th=i\pi/3}S_0(\th)=
\nonumber\\
&& \mbox{Res}_{\th=2i\pi/3}S_1(\th)=-\mbox{Res}_{\th=i\pi/3}S_2(\th)=
i(\Gamma_{KK}^K)^2\,, \nonumber
\eea
determine the coupling at the three-kink vertex (Fig.\,2a)
\EQ
\Gamma_{KK}^{K}=\left[\frac{1}{\lambda}\sin\frac{2\pi\lambda}{3}\,
e^{{\cal A}(i\frac{\pi}{3})}\right]^{1/2}\,\,.
\EN
For $q>3$ ($\lambda>1$) a direct channel (positive residue) pole located at
$\th=2i\kappa$,
\EQ
\kappa=\frac{\pi}{2}\left(1-\frac{1}{\lambda}\right)\,,
\EN
enters the physical strip in the amplitudes $S_2(\th)$ and $S_3(\th)$. Such
a pole must be accordingly associated to a (topologically neutral)
kink-antikink bound state $B$ with mass 
\EQ
m_B=2m\cos\kappa\,\,.
\EN
Of course, the amplitudes $S_1(\th)$ and $S_3(\th)$ exhibit the corresponding 
crossed channel (negative residue) pole at $\th=i\pi-2i\kappa$. The coupling
at the kink-kink-bound state vertex (Fig.\,2b) is given by
\EQ
\Gamma_{KK}^B=\left[-i\mbox{Res}_{\th=2i\kappa}S_3(\th)\right]^{1/2}=
\left[\frac{1}{\lambda}\sin\frac{4\pi\lambda}{3}\,\frac{\sin\lambda\pi}
{\sin\frac{\lambda\pi}{3}}\,e^{{\cal A}(2i\kappa)}\right]^{1/2}\,\,.
\label{fb}
\EN
The amplitudes $S_{KB}(\th)$ and $S_{BB}(\th)$ (Fig.\,1b,c) describing the 
kink-bound  state scattering and the bound state self-interaction are 
determined by the bootstrap equations
\bea
&& S_{KB}(\th)=(q-2)S_2(\th-i\kappa)S_1(\th+i\kappa)+
S_3(\th-i\kappa)S_3(\th+i\kappa)\,,\nonumber\\
&& S_{BB}(\th)=S_{BK}(\th-i\kappa)S_{BK}(\th+i\kappa)\,,
\eea
and read
\bea
&& S_{BK}(\th)=t_{1-\kappa/\pi}(\th)t_{2/3-\kappa/\pi}(\th)\,,\nonumber\\
&& S_{BB}(\th)=t_{2/3}(\th)t_{1-2\kappa/\pi}(\th)t_{2/3-2\kappa/\pi}(\th)\,,
\eea
in terms of the building blocks
\EQ
t_a(\th)=\frac{\tanh\frac{1}{2}(\th+i\pi a)}{\tanh\frac{1}{2}(\th-i\pi a)}\,\,.
\EN
The poles located at $\th=i(\pi-\kappa)$ in $S_{BK}$ and at $\th=2\pi/3$ 
in $S_{BB}$ are bound state poles corresponding to $K$ and $B$, respectively.
The coupling at the $BBB$ vertex (Fig.\,2c) is
\bea
&& \Gamma_{BB}^B=\left[-i\,\mbox{Res}_{\th=2i\pi/3}S_{BB}(\th)\right]^{1/2}=
\nonumber\\
&& \left[2\sqrt{3}\cot\frac{\pi}{2}\left(1-\frac{1}{\lambda}\right)
\cot\pi\left(\frac{1}{2\lambda}-\frac{1}{3}\right)
\tan\frac{\pi}{6}\left(4-\frac{3}{\lambda}\right)
\tan\frac{\pi}{6}\left(5-\frac{3}{\lambda}\right)\right]^{1/2}.
\eea
It can be shown \cite{patrick} that, at least in the region $q\leq 4$ we are 
interested in, the remaining poles in the amplitudes $S_{KB}$ and $S_{BB}$ are 
associated to multi-scattering processes
rather than to new particles\footnote{Multi-scattering singularities usually
show up as higher order poles in two dimensions (they lead to anomalous
thresholds in four dimensions) \cite{CT}. In the present case a simultaneous 
vanishing of the residues reduces the singularities to simple poles 
\cite{patrick}.}. Hence, the elementary kinks and their neutral bound state 
$B$ are the only particles entering the spectrum of the theory in this range 
of the parameter $q$.

\resection{Form factors}
The two-kink form factor of an operator $\Phi(x)$ (Fig.\,3) is defined as the 
matrix element\footnote{We will consider only operators which are scalar under 
Lorentz transformations. Their matrix elements depend on rapidity differences 
only.}
\EQ
F^{\Phi}_{\alpha\gamma\beta}(\th_{12})\equiv\langle 0_\al|\Phi(0)|
K_{\alpha\gamma}(\th_1)K_{\gamma\beta}(\th_2)\rangle\,,
\label{ff}
\EN
where $|0_\al\rangle$ denotes the vacuum state in which all the sites have 
colour $\al$.
The fundamental equations constraining this matrix element come from the 
requirements of unitarity and crossing symmetry. The unitarity equation for
form factors follows immediately from Eq.\,(\ref{fz}) 
\EQ
F^{\Phi}_{\alpha\gamma\beta}(\th)=\sum_{\delta\neq\alpha,\beta}
S_{\al\beta}^{\gamma\delta}(\th)F^{\Phi}_{\alpha\delta\beta}(-\th)\,\,.
\label{ffunitarity}
\EN
The crossing equations read (see the Appendix)
\EQ
F^{\Phi}_{\alpha\gamma\al}(\th+2i\pi)=F^{\Phi}_{\gamma\al\gamma}(-\th)\,,
\label{ffcrossing1}
\EN
for the neutral sector, and 
\EQ
F^{\Phi}_{\alpha\gamma\beta}(\th+2i\pi)=F^{\Phi}_{\al\gamma\beta}(-\th)\,,
\hspace{.5cm}\al\neq\beta
\label{ffcrossing2}
\EN
for the charged sector. Let us consider how 
Eqs.\,(\ref{ffunitarity})--(\ref{ffcrossing2}) specialise for the different 
operators we are interested in.

\subsection{Energy operator}
The energy operator $\varepsilon(x)$ is the operator which perturbs conformal
invariance in the action (\ref{action}). It is then proportional to the 
trace of the stress-energy tensor $\Theta(x)$
\EQ
\Theta(x)=2\pi (2-x_\varepsilon)\tau\,\varepsilon(x)\,\,.
\label{thetaeps}
\EN
In view of this relation, we will mainly refer to $\Theta(x)$ in the following.

Being invariant under the $S_q$ symmetry, $\Theta(x)$ couples to the neutral 
sector of the space of states. Moreover, its two-kink form factors
\EQ
F^{\Theta}(\th_{12})\equiv\langle 0_\al|\Theta(0)|
K_{\alpha\gamma}(\th_1)K_{\gamma\al}(\th_2)\rangle\,,
\label{traceff}
\EN
do not depend on the choice of $\al$ and $\gamma$. The unitarity and crossing
equations can then be written in the form
\bea
&& F^\Theta(\th)=\Lambda(\th)F^\Theta(-\theta)\,,\nonumber\\
&& F^\Theta(\th+2i\pi)=F^\Theta(-\theta)\,,
\label{thetasystem}
\eea
where
\EQ
\Lambda(\th)=(q-2)S_2(\th)+S_3(\th)=\frac{\sinh\lambda(\th+i\pi)}
{\sinh\lambda(\th-i\pi)}\,{\cal E}(\th)\,,
\EN
\EQ
{\cal E}(\th)=\exp\left[\int_0^\infty\frac{dx}{x}\,g_{{\cal E}}(x)
\sinh\frac{x\th}{i\pi}\right]\,,
\EN
\EQ
g_{{\cal E}}(x)=2\,\frac{\sinh\frac{x}{3}\cosh\left(\frac{1}{3}-
\frac{1}{\lambda}\right)\frac{x}{2}}{\sinh\frac{x}{2\lambda}
\cosh\frac{x}{2}}\,\,.
\EN
Since
\EQ
\frac{\sinh\lambda(\th+i\pi a)}{\sinh\lambda(\th-i\pi a)}=
-\exp\left[\int_0^\infty\frac{dx}{x}\,g_a(x)\sinh\frac{x\th}{i\pi}\right]\,,
\EN
\EQ
g_a(x)=2\,\frac{\sinh\left(\frac{1}{2\lambda}-a\right)x}
{\sinh\frac{x}{2\lambda}}\,,
\EN
it is easily checked that the function
\EQ
F_{\Lambda}(\th)=-i\sinh\frac{\th}{2}\,\exp\left\{\int_0^\infty\frac{dx}{x}\,
\frac{g_1(x)+g_{{\cal E}}(x)}{\sinh x}\,\sin^2\frac{(i\pi-\th)x}{2\pi}
\right\}\,,\hspace{.5cm}\lambda<1
\label{flambda}
\EN
solves the system of functional equations (\ref{thetasystem}). We note here
for later convenience the asymptotic behaviour
\EQ
F_\Lambda(\th)\sim\exp\left(1-\frac{2\lambda}{3}\right)\th\,,\hspace{1cm}
\th\goto +\infty\,\,.
\label{lambdasymp}
\EN
The integral in (\ref{flambda}) is convergent on the real $\th$-axis for 
$\lambda<1$. For $\lambda> 1$ one needs to use the analytically continued 
expression
\EQ
F_\Lambda(\th)=\frac{-\cos^2\kappa}
{\sinh\frac{1}{2}(\th-2i\kappa)\sinh\frac{1}{2}(\th+2i\kappa)}\,
\Upsilon(\th)\,,
\hspace{.5cm}\lambda>1
\EN
\EQ
\Upsilon(\th)=-i\sinh\frac{\th}{2}\,\exp\left\{\int_0^\infty\frac{dx}{x}\,
\frac{g_{1-1/\lambda}(x)+g_{{\cal E}}(x)}{\sinh x}\,
\sin^2\frac{(i\pi-\th)x}{2\pi}\right\}\,,
\EN
which explicitly exhibits the pole at $\th=2i\kappa$ in the physical strip 
corresponding to the kink-antikink bound state $B$ (Fig.\,4a).
Taking into account the normalisation condition
\EQ
F^\Theta(i\pi)=2\pi m^2\,,
\EN
we can write
\EQ
F^{\Theta}(\th)=2\pi m^2 F_{\Lambda}(\th)\,,
\EN
\EQ
F_B^\Theta\equiv\langle 0_\al|\Theta(0)|B\rangle=\frac{1}{i\Gamma_{KK}^B}
\mbox{Res}_{\th=2i\kappa}F^\Theta(\th)=
\frac{\pi m^2_B}{\sin 2\kappa}\,\frac{\Upsilon(2i\kappa)}{\Gamma_{KK}^B}\,\,.
\EN
Let us finally determine the matrix element $F_{BB}^\Theta(\th_{12})\equiv
\langle 0_\al|\Theta(0)|B(\th_1)B(\th_2)\rangle$. It has to satisfy the  
monodromy equations
\bea
&& F_{BB}^\Theta(\th)=S_{BB}(\th)F_{BB}^\Theta(-\th)\,,\nonumber\\
&& F_{BB}^\Theta(\th+2i\pi)=F_{BB}^\Theta(-\th)\,\,.
\eea
The solution to these equations with the expected pole structure is
\EQ
F_{BB}^\Theta(\th)=[a\cosh\th+b]\sinh\frac{\th}{2}\,
{\cal R}_{2/3}(\th){\cal R}_{1-2\kappa/\pi}(\th)
{\cal R}_{2/3-2\kappa/\pi}(\th)\,,
\label{fbb}
\EN
where
\EQ
{\cal R}_{a}(\th)=\frac{1}{\cosh\th-\cos\pi a}
\exp\left\{2\int_0^\infty\frac{dx}{x}\,
\frac{\cosh\left(a-\frac{1}{2}\right)x}{\cosh\frac{x}{2}\sinh x}\,
\sin^2\frac{(i\pi-\th)x}{2\pi}\right\}\,.
\EN
The coefficients $a$ and $b$ in (\ref{fbb}) are uniquely determined by
the residue equation
\EQ
-i\,\mbox{Res}_{\th=2i\pi/3}F_{BB}^\Theta(\th)=\Gamma_{BB}^BF_B^\Theta\,,
\EN
and the normalisation condition
\EQ
F_{BB}^\Theta(i\pi)=2\pi m_B^2\,\,.
\EN
Having computed $F_B^\Theta$ and $F_{BB}^\Theta(\th)$, we can determine the
vacuum expectation value $F_0^\Theta\equiv\langle 0_\al|\Theta|0_\al\rangle$
through the asymptotic factorisation relation \cite{DSC}
\EQ
\lim_{\th\goto\infty}F_{BB}^\Theta(\th)=\frac{(F_B^\Theta)^2}
{F_0^\Theta}\,\,.
\EN
We find
\EQ
F_0^\Theta=-\frac{\pi\sin\frac{\pi}{2\lambda}}
{\sqrt{3}\sin\pi\left(\frac{1}{3}+\frac{1}{2\lambda}\right)}\,m^2\,,
\label{vevtrace}
\EN
in perfect agreement with the result obtained in \cite{Fateev} using the 
thermodynamic Bethe ansatz (TBA).

\subsection{Disorder operators}
The theory contains disorder operators which are dual to the $q-1$
independent order operators $\sigma_\al(x)$ and have the same scaling 
dimension (\ref{xsigma}). Following the usual 
interpretation of the disorder operator as the kink creation operator,
we denote by $\mu_\beta(x)$ the disorder operator which, acting on the 
vacuum state $|0_\alpha\rangle$, creates states interpolating between
the vacua $\alpha$ and $\beta$ ($\beta\neq\alpha$). The two-kink form factor
\EQ
\langle 0_\al|\mu_\beta(0)|K_{\alpha\gamma}(\th_1)
K_{\gamma\delta}(\th_2)\rangle=\delta_{\beta\delta}F^\mu(\th_{12})\,,
\hspace{.5cm}\beta\neq\al
\label{muff}
\EN
does not depend on the intermediate index $\gamma$ and satisfies the unitarity 
and crossing equations
\bea
&& F^\mu(\th)=\Sigma(\th)F^\mu(-\theta)\,,\nonumber\\
&& F^\mu(\th+2i\pi)=F^\mu(-\theta)\,,
\label{musystem}
\eea
with
\EQ
\Sigma(\th)=(q-3)S_0(\th)+S_1(\th)=
\frac{\sinh\lambda\left(\th+\frac{2i\pi}{3}\right)}
{\sinh\lambda\left(\th-\frac{2i\pi}{3}\right)}\,{\cal E}(\th)\,\,.
\EN
Hence, it can be written as
\EQ
F^\mu(\th)=Z_\mu\,\frac{F_\Sigma(\th)}{\sinh\frac{1}{2}
\left(\th+\frac{2i\pi}{3}\right)\sinh\frac{1}{2}
\left(\th-\frac{2i\pi}{3}\right)}\,,
\EN
where
\EQ
F_{\Sigma}(\th)=-i\sinh\frac{\th}{2}\,\exp\left\{\int_0^\infty\frac{dx}{x}\,
\frac{g_{2/3}(x)+g_{{\cal E}}(x)}{\sinh x}\,\sin^2\frac{(i\pi-\th)x}{2\pi}
\right\}\,,
\EN
is the solution of the system (\ref{musystem}) without poles in the physical
strip; the pole at $\th=2i\pi/3$ corresponding to the $\varphi^3$-property
of the kinks has been inserted explicitly. The function $F_\Sigma(\th)$ 
behaves asymptotically as
\EQ
F_\Sigma(\th)\sim\exp\left(1-\frac{\lambda}{3}\right)\th\,,\hspace{1cm}
\th\goto +\infty\,\,.
\EN
The one-kink form factor
\EQ
\langle 0_\al|\mu_\beta(0)|K_{\al\gamma}\rangle=\delta_{\beta\gamma}F^\mu_K\,,
\hspace{.5cm}\beta\neq\al
\EN
is given by (Fig.\,4b)
\EQ
F^\mu_K=\frac{1}{i\Gamma_{KK}^K}\mbox{Res}_{\th=2i\pi/3}F^\mu(\th)=
-\frac{4F_\Sigma(2i\pi/3)}{\sqrt{3}\,\Gamma_{KK}^K}\,Z_\mu\,\,.
\EN

\subsection{Magnetisation operators}
The magnetisation operators $\sigma_\al(x)$ couple to neutral states. 
Since $\sigma_\al$ carries an index, its matrix
elements are not in general invariant under permutations. In the two-kink form
factor we distinguish the three components
\EQ
\langle 0_\al|\sigma_\beta(0)|K_{\al\gamma}(\th_1)K_{\gamma\al}(\th_2)\rangle=
\delta_{\beta\al}F^\sigma_1(\th_{12})+\delta_{\beta\gamma}F^\sigma_2(\th_{12})
+(1-\delta_{\beta\alpha}-\delta_{\beta\gamma})F^\sigma_3(\th_{12})\,,
\EN
which eq.\,(\ref{constraint}) relates as
\EQ
F^\sigma_1(\th)+F^\sigma_2(\th)+(q-2)F^\sigma_3(\th)=0\,\,.
\label{ffconstraint}
\EN
Eq.\,(\ref{ffunitarity}) provides the relations
\bea
&& F^\sigma_1(\th)=\Lambda(\th)F_1^\sigma(-\th)\,,\nonumber\\
&& F^\sigma_2(\th)=S_3(\th)F_2(-\th)+(q-2)S_2(\th)F^\sigma_3(-\th)\,,
\nonumber\\
&& F^\sigma_3(\th)=S_2(\th)F^\sigma_2(-\th)+[(q-3)S_2(\th)+S_3(\th)]
F^\sigma_3(-\th)\,\,.
\label{sigmaunitarity}
\eea
The crossing equations
\bea
&& F^\sigma_1(\th+2i\pi)=F^\sigma_2(-\th)\,,\nonumber\\
&& F^\sigma_3(\th+2i\pi)=F^\sigma_3(-\th)\,,
\label{sigmacrossing}
\eea
are an immediate consequence of (\ref{ffcrossing1}).
Recalling that the function $F_\Lambda(\th)$ defined in (\ref{flambda}) 
satisfies Eqs.\,(\ref{thetasystem}), we parameterise $F^\sigma_1(\th)$ as
\EQ
F^\sigma_1(\th)=\Omega(\th)F_\Lambda(\th)\,;
\label{fsig1}
\EN
then, the first of (\ref{sigmaunitarity}) is automatically fulfilled provided
the function $\Omega(\th)$ satisfies
\EQ
\Omega(\th)=\Omega(-\th)\,\,.
\label{even}
\EN
Simple manipulations involving Eqs.\,(\ref{ffconstraint})-(\ref{sigmacrossing})
provide the basic equation that $\Omega(\th)$ should satisfy together
with the previous one:
\EQ
\Omega(\th)=\left[\frac{S_3(\th)}{S_2(\th)}-1\right]\Omega(2i\pi+\th)-
\frac{\Lambda(\th)}{S_2(\th)}\,\Omega(2i\pi-\th)\,,
\EN
or, more explicitly,
\EQ
\Omega(\th)=\frac{\sinh\lambda(\th-i\pi)\,\Omega(2i\pi+\th)+
\sinh\lambda(\th+i\pi)\,\Omega(2i\pi-\th)}
{2\cos\frac{\pi\lambda}{3}\,\sinh\lambda\th}\,\,.
\label{omega}
\EN
Unfortunately, we do not know how to solve this equation
for generic values of $\lambda$. Before turning to the solution
for specific values of this parameter, we collect some additional physical 
information about the magnetisation matrix elements.

The order operators $\sigma_\al(x)$ are non-local with respect to the 
disorder operators $\mu_\beta(x)$ which interpolate the kinks. This kind of 
mutual non-locality is known to lead to the presence of an ``annihilation
pole'' at $\theta=i\pi$ in the two-particle form factor \cite{Smirnov,YZ,msg}.
We show in the Appendix that in our case the residue on such a pole takes the 
form
\EQ
-i\,\mbox{Res}_{\th_1-\th_2=i\pi}\langle 0_\al|\sigma_\gamma(0)|
K_{\al\beta}(\th_1)K_{\beta\al}(\th_2)\rangle=
\langle 0_\al|\sigma_\gamma|0_\al\rangle-
\langle 0_\beta|\sigma_\gamma|0_\beta\rangle\,\,.
\EN
Denoting
\EQ
F_0^\sigma\equiv \langle 0_\al|\sigma_\al|0_\al\rangle\,,
\EN
and taking into account the constraint (\ref{constraint}), the order 
parameter can be written as
\EQ
\langle 0_\al|\sigma_\gamma|0_\al\rangle=
\frac{F_0^\sigma}{q-1}\,(q\delta_{\gamma\al}-1)\,\,.
\label{vev}
\EN
Hence, for the different components of the two-kink form factor we conclude
\bea
&& -i\,\mbox{Res}_{\th=i\pi}F_1^\sigma(\th)=
-i\,\mbox{Res}_{\th=i\pi}F_2^\sigma(\th)=\frac{q}{q-1}\,F_0^\sigma \nonumber\\
&& -i\,\mbox{Res}_{\th=i\pi}F_3^\sigma(\th)=0
\label{annihilation}
\eea

For $\lambda>1$ the magnetisation has a one-particle form factor on the 
bound state $B$. Denoting
\EQ
F_B^\sigma\equiv\langle 0_\al|\sigma_\al(0)|B\rangle\,,
\EN
we can write
\EQ
\langle 0_\al|\sigma_\gamma(0)|B\rangle=\frac{F_B^\sigma}{q-1}\,
(q\delta_{\gamma\al}-1)\,\,.
\label{fsigb}
\EN
These one-particle form factors can be obtained from the two-kink matrix 
elements through the residue equation
\EQ
-i\,\mbox{Res}_{\th_1-\th_2=2i\kappa}\langle 0_\al|\sigma_\gamma(0)|
K_{\al\beta}(\th_1)K_{\beta\al}(\th_2)\rangle=
\Gamma_{KK}^B\,\langle 0_\al|\sigma_\gamma(0)|B\rangle\,\,.
\EN
The last two equations, together with the crossing relation 
(\ref{sigmacrossing}), immediately lead to the following identity for the 
function $\Omega(\th)$ introduced in (\ref{fsig1})
\EQ
\Omega(2i\pi-2i\kappa)=-\frac{1}{q-1}\,\Omega(2i\kappa)\,\,.
\label{identity}
\EN
Although obtained for $\lambda>1$, this relation is expected to hold for 
generic values of the parameter.

The matrix elements of scaling operators in unitary theories must satisfy the 
asymptotic bound obtained in Ref.\,\cite{immf}. For the kink-kink form factor 
of the magnetisation operators this reads
\EQ
\lim_{\th\goto +\infty}F_1^\sigma(\th)\leq constant\,e^{x_\sigma\th/2}\,\,.
\label{bound}
\EN

The functional equation (\ref{omega}) becomes trivial at the points
$q=2,3,4$. Let us investigate these cases in more detail.

\noindent
{\bf q=2}

\noindent
For this value of $q$ (corresponding to $\lambda=3/4$) the solution to 
Eqs.\,(\ref{even}) and (\ref{omega}) satisfying the asymptotic bound 
(\ref{bound}) is simply 
\EQ
\Omega(\th)=\frac{iF_0^\sigma}{\cosh\frac{\th}{2}}\,,
\EN
where the normalisation has been fixed using Eq.\,(\ref{annihilation}).
The identity (\ref{identity}) is automatically satisfied. Since
$F_\Lambda(\th)=-i\sinh\th/2$, one obtains
\EQ
F_1^\sigma(\th)=-F_2^\sigma(\th)=iF_0^\sigma\tanh\frac{\th}{2}\,,
\EN
which is the well known result usually obtained in the high temperature
formalism \cite{BKW,YZ}.\vspace{.6cm}

\noindent
{\bf q=3}

\noindent
For $\lambda=1$ Eq.\,(\ref{omega}) reduces to 
\EQ
\Omega(\th)=-[\Omega(\th+2i\pi)+\Omega(\th-2i\pi)]\,\,.
\EN
Together with Eqs.\,(\ref{even}), (\ref{bound}) and (\ref{annihilation}), 
it fixes\footnote{An equivalent result can be obtained working with the high 
temperature scattering theory which is defined in terms of particles rather
than kinks \cite{3Potts,inrussian}.}
\EQ
\Omega(\th)=-\frac{\sqrt{3}}{2}\,F_0^\sigma\,\frac{\cosh\frac{\th}{6}}
{\cosh\frac{\th}{2}}\,\,.
\EN
Again Eq.\,(\ref{identity}) is automatically satisfied.\vspace{.6cm}

\noindent
{\bf q=4}

\noindent
For $\lambda=3/2$ Eq.\,(\ref{omega}) becomes
\EQ
\Omega(\th+2i\pi)=\Omega(\th-2i\pi)\,\,.
\EN
This time Eq.\,(\ref{identity}) has to be enforced together with the other 
constraints in order to fix the solution
\EQ
\Omega(\th)=-\frac{2F_0^\sigma}{3\sqrt{3}}\,\frac{\cosh\frac{\th}{2}+\sqrt{3}}
{\cosh\frac{\th}{2}}\,\,.
\EN

We conclude this section considering the problem of the relative normalisation
of the order and disorder operators we discussed in the introduction.
Solving Eq.\,(\ref{omega}) for $\th\goto\infty$ one easily finds that 
$\Omega(\th)$ has to behave in this asymptotic limit as $e^{\eta\th}$, with
$\eta=\lambda/3+k$ {\em or} $\eta=2\lambda/3+k$ ($k$ integer). The solutions
obtained at $q=2,3,4$ indicate that the correct choice is $\eta=2\lambda/3-1$.
Combining this result with the asymptotic behaviour (\ref{lambdasymp}) for 
$F_\Lambda(\th)$, we conclude that $F_1^\sigma(\th)$ goes to a constant in the
limit of large rapidity difference. Hence, according to the discussion of 
Ref.\,\cite{DSC}, we expect the following asymptotic factorisation equation
to hold
\EQ
\lim_{\th\goto\infty}|F_1^\sigma(\th)|=\frac{(F_K^\mu)^2}{F_0^\sigma}\,\,.
\EN
In the next section, this relation will enable us to express the critical 
amplitudes in terms of a single arbitrary normalisation constant (say
$F_0^\sigma$) which in turn cancels out when the universal amplitude ratios 
are considered.

\resection{Correlation functions and amplitude ratios}
The knowledge of the form factors allows to express correlation functions 
as spectral sums over intermediate asymptotic states. For the two-point
euclidean correlator of two scalar operators $\Phi_1(x)$ and $\Phi_2(x)$ 
we have
\EQ
\langle\Phi_1(x)\Phi_2(0)\rangle=\sum_{n=0}^{\infty}
\int_{\th_1>\ldots>\th_n}\frac{d\th_1}{2\pi}\ldots\frac{d\th_n}{2\pi}\,
\langle 0|\Phi_1(0)|n\rangle\langle n|\Phi_2(0)|0\rangle e^{-|x|E_n}\,,
\label{spectral}
\EN
where $E_n$ denotes the total energy of the $n$-particle state $|n\rangle$.

Clearly, the above expression is a large distance expansion: while 
the intermediate states with the lowest total mass provide the dominant
contribution for large separations, in principle the whole series should be 
resummed in order to reproduce the correct ultraviolet behaviour.
Relying on the properties of fast convergence of the spectral series we 
mentioned in the introduction,
we consider partial sums of the series truncated at 
the level of the two-kink intermediate state (the relevant form factors have 
been computed in the previous section). For example, the expansions for the 
two-point correlators of the order and disorder parameters read
\bea
\langle 0_\al|\sigma_\beta(x)\sigma_\gamma(0)|0_\alpha\rangle &=&
\frac{(q\delta_{\beta\al}-1)(q\delta_{\gamma\al}-1)}{(q-1)^2}
\left[|F_0^\sigma|^2+H(q-3)\frac{|F_B^\sigma|^2}{\pi}K_0(m_B|x|)\right]+
\nonumber\\
&+&\frac{1}{\pi^2}\int_0^\infty d\th\,f_{\beta\gamma}^\al(2\th)\,
K_0(2m|x|\cosh\th)\,+\ldots\label{sigma-sigma}\\
\langle 0_\al|\mu_\beta(x)\mu_\gamma(0)|0_\al\rangle &=&
\delta_{\beta\gamma}\left[\frac{|F_k^\mu|^2}{\pi}K_0(m|x|)\,+\right.\nonumber\\
&+& \left.\frac{q-2}{\pi^2}\int_0^\infty d\th\,|F^\mu(2\th)|^2\,
K_0(2m|x|\cosh\th)\right]+\ldots\hspace{.4cm}\beta,\gamma\neq\al
\label{mu-mu}
\eea
where we introduced the function
\bea
f_{\beta\gamma}^\al(\th)&=&[\delta_{\beta\al}\delta_{\gamma\al}(q-1)
-\delta_{\beta\al}(1-\delta_{\gamma\al})-(1-\delta_{\beta\al})
\delta_{\gamma\al}]|F_1^\sigma(\th)|^2\,+\nonumber\\
&+&\delta_{\beta\gamma}(1-\delta_{\beta\al})[|F_2^\sigma(\th)|^2+
(q-2)|F_3^\sigma(\th)|^2]\,+\nonumber\\
&+&(1-\delta_{\beta\gamma})(1-\delta_{\beta\al})
[F_2^\sigma(\th)F_3^\sigma(-\th)+F_3^\sigma(\th)F_2^\sigma(-\th)+
(q-3)|F_3^\sigma(\th)|^2]\,,
\eea
and the step function $H(y)$ which equals $1$ for $y>0$ and is $0$ 
otherwise.
The terms we omitted in the r.h.s. are of order $e^{-3m|x|}$ for $|x|$ large 
(at least for $q\leq 3$). We show immediately that this level of approximation
is sufficient to obtain remarkably accurate numerical results for the 
quantities we need to evaluate in this paper, namely moments of two-point
correlators of the type 
$\int d^2x|x|^p\langle\Phi_1(x)\Phi_2(0)\rangle_c$. An interesting check is 
provided by the following sum rules for the central charge of the ultraviolet
CFT \cite{cth} and the scaling dimension of the magnetisation operator
\cite{DSC}
\bea
&& c=\frac{3}{4\pi}\int d^2x\,|x|^2\langle 0_\al|\Theta(x)\Theta(0)|0_\al
\rangle_c\,,\nonumber\\
&& x_\sigma=-\frac{1}{2\pi\langle 0_\al|\sigma_\gamma|0_\al\rangle}
\int d^2x\,\langle 0_\al|\Theta(x)\sigma_\gamma(0)|0_\al\rangle_c\,,
\label{sumrules}
\eea
where $\langle\cdots\rangle_c$ denotes connected correlators. The result 
obtained for $c$ as a function of $q$ using the truncated spectral expansion 
of the trace-trace correlator is shown in Fig.\,5 and compared with the 
exact value (\ref{c})\footnote{It can be appreciated from the figure that the 
values of $c$ obtained from this first contribution to the spectral sum are
slightly larger than the exact result in the range $1<q<2$. At first sight
this may seem strange since each intermediate state which remains to be 
included into the sum will give a positive contribution (an integral over the 
modulus square of the corresponding trace form factor). Notice however that 
the number of three-kink intermediate states in the neutral topological
sector is $(q-1)(q-2)$ and becomes negative when $1<q<2$!}. The numerical 
results obtained for $c$ and $\Delta_\sigma$ for $q=2,3,4$ are listed in 
Table\,1. Since the energy operator in the two-dimensional Ising model couples 
to the two-kink state only, the results obtained for $q=2$ are exact. In the 
other cases the observed deviation from the exact result is of few percent 
at most.

The critical amplitudes (\ref{amplitudes}) are linked to the 
off-critical correlators by the relations
\EQ
C=\int d^2x\,\langle 0_\al|\varepsilon(x)\varepsilon(0)|0_\al\rangle_c\,\,,
\label{heat}
\EN
\EQ
M=\langle 0_\al|\sigma_\al|0_\al\rangle\,,
\label{mag}
\EN
\EQ
\chi=\int d^2x\,\langle 0_\al|\sigma_\al(x)\sigma_\al(0)|0_\al\rangle_c\,\,,
\label{chi}
\EN
\EQ
\xi^2=\frac{1}{4}\,
\frac{\int d^2x\,|x|^2\langle 0_\al|\sigma_\al(x)\sigma_\al(0)|0_\al\rangle_c}
{\int d^2x\,\langle 0_\al|\sigma_\al(x)\sigma_\al(0)|0_\al\rangle_c}\,\,.
\label{csi}
\EN
The dimensional parameter entering our $S$-matrix approach is the mass $m$
of the kink. It is related to the reduced temperature $\tau$ appearing
in (\ref{amplitudes}) as\footnote{The dimensionless number $m_0$ is exactly
computable through the TBA \cite{Fateev}; we will not need it here.} 
\EQ
m=m_0\,\tau^\nu\,\,.
\EN
Equation (\ref{csi}) defines the ``second moment'' correlation length. In the 
literature the so-called ``true'' correlation length $\xi_t$ is often 
considered which is defined through the large distance asymptotic decay of the 
spin-spin correlators
\EQ
\langle\sigma(x)\sigma(0)\rangle\sim \exp(-|x|/\xi_t)\,,\hspace{1cm}
|x|\goto\infty\,\,.
\EN
It follows from (\ref{mu-mu}) together with duality that $\xi_t=1/m$ at 
$T>T_c$. At $T<T_c$, Eq.\,(\ref{sigma-sigma}) implies instead $\xi_t=1/2m$
for $q\leq 3$, and $\xi_t=1/m_B$ for $3<q\leq 4$.

Since the zeroth moment of the energy-energy correlator appears in 
the scaling dimension sum rule
\EQ
x_\varepsilon=-\frac{1}{2\pi\langle 0_\al|\varepsilon|0_\al\rangle}
\int d^2x\,\langle 0_\al|\Theta(x)\varepsilon(0)|0_\al\rangle_c\,,
\EN
the specific heat amplitudes $A_\pm$ can be computed exactly as\footnote{The 
specific heat diverges logarithmically 
in the Ising model ($\al=0$) and the definition of the amplitudes is 
accordingly modified to $C\simeq-A_\pm\ln\tau$. In the limit $q\goto 2$, 
Eq.\,(\ref{a}) gives the correct result $A_\pm=m_0^2/2\pi$.}
\EQ
A_\pm=-\al(1-\al)(2-\al)\left(\frac{m_0}{m}\right)^2\frac{F_0^\Theta}{4\pi}\,,
\label{a}
\EN
where $F_0^\Theta$ is given in (\ref{vevtrace}).
The equality of $A_+$ and $A_-$ follows from the fact that the energy operator
$\varepsilon(x)$ simply changes sign under the duality transformation
exchanging the low and high temperature phases.

Remembering the definition (\ref{vev}), the magnetisation amplitude $B$ can
be written as
\EQ
B=\left(\frac{m_0}{m}\right)^{x_\sigma}F_\sigma^0\,\,.
\label{b}
\EN
The susceptibility and correlation length amplitudes can be evaluated using
the expansions (\ref{sigma-sigma}) and (\ref{mu-mu}) for the correlators and 
the corresponding form factors. By duality, the
high temperature amplitudes are obtained substituting $\sigma_\al$ by
$\mu_\beta$ ($\beta\neq\al$) in Eqs.\, (\ref{chi}) and (\ref{csi}). Some of 
the values obtained in this way for integer $q$ are listed in Table\,2. In 
particular, the results for $\xi_0^+m_0$ show that the ``second moment'' and 
``true'' correlation lengths differ very slightly from each other at $T>T_c$.

Table\,3 collects the results we find for the amplitude 
ratios at $q=2,3,4$. Excepting $A_+/A_-$, we are not aware of previous reliable
estimates of these ratios for the cases $q=3,4$. At $q=2$, however, comparison
with the known exact results \cite{McW,WMcTB} (see also \cite{ratios}) 
\bea
&& \Gamma_+/\Gamma_-=37.69365..\,,\nonumber\\
&& R_C=0.318569..\,,\nonumber\\
&& \xi_0^+/\xi_0^-=3.16..\,,
\eea
further confirms the remarkable accuracy of the results yielded by our 
two-kink approximation.

\resection{Percolation}
Percolation is the purely geometrical problem (no temperature involved) in 
which bonds\footnote{We refer here to {\em bond} percolation; in {\em site} 
percolation, sites rather than bonds are occupied with probability $p$. Of 
course, all universal results are independent of this distinction.}  are 
randomly distributed on a lattice with occupation probability $p$ \cite{SA}. 
A set of bonds forming a connected path on the lattice is called a {\em 
cluster}. There exist a critical value $p_c$ of the occupation probability 
above which an infinite cluster appears in the system; $p_c$ is called the 
{\em percolation threshold}. If $N$ is the total number of bonds in the 
lattice, the probability of a configuration with $N_b$ occupied bonds is 
$p^{N_b}(1-p)^{N-N_b}$. Hence, the average of a quantity $X$ over all 
configurations ${\cal G}$ is 
\EQ
\langle X\rangle=\sum_{\cal G}X\,p^{N_b}(1-p)^{N-N_b}\,\,.
\label{average}
\EN
Let $x$ and $y$ denote the positions of two bonds on the lattice and consider
the function $C(x,y)$ which takes value $1$ if $x$ and $y$ belong to the 
same cluster, and $0$ otherwise. Then, the function $g(x,y)=
\langle C(x,y)\rangle$ is the probability that $x$ and $y$ belong to the same
cluster and is called {\em pair connectivity}. If we denote by $P$ the 
probability that a bond belongs to the infinite cluster ($P=0$ at $p<p_c$), 
clearly we have
\EQ
\lim_{|x-y|\goto\infty}g(x,y)=P^2\,\,.
\label{p}
\EN
Since $\sum_y C(x,y)$ counts the total number of bonds in the cluster $x$ 
belongs to, the {\em mean cluster size} can be obtained as
\EQ
S=\sum_y g_c(0,y)\,,
\label{size}
\EN
where the subscript $c$ means that the connected part of the pair connectivity
is taken in order to get rid of the contribution coming from the infinite 
cluster at $p>p_c$. The second moment correlation length is also naturally
defined in terms of the pair connectivity as
\EQ
\tilde{\xi}^2=\frac{1}{2d}\,\frac{\sum_x|x|^2g_c(x,0)}{\sum_xg_c(x,0)}\,\,.
\label{xitilde}
\EN

The following relation with the $q$-state Potts model is particularly 
important for the theoretical study of percolation processes \cite{KF}. 
Remembering (\ref{newpartition}), the average of a quantity
$X$ in the $q$-state Potts model can be written as
\EQ
\langle X\rangle_q=Z^{-1}\sum_{\cal G}X\,q^{N_c}x^{N_b}\,;
\EN
hence, it is sufficient to make the formal identification $x=p/(1-p)$ to see
that $\langle X\rangle_1$ coincides with the percolation average 
(\ref{average}). For example, the {\em mean cluster number} in percolation
can be expressed as
\EQ
\langle N_c\rangle=\lim_{q\goto 1}\frac{\partial\ln Z}{\partial q}=
\lim_{q\goto 1}\frac{\ln Z}{q-1}\,\,.
\label{number}
\EN

To proceed further with this mapping we need a representation of pair 
connectivity in the Potts model formalism. For this purpose observe that
the insertion of a delta function $\delta_{s(x)\al}$ in a Potts configuration
fixes to $\al$ the colour of the cluster $x$ belongs to. Hence, at $T>T_c$,
\bea
&& \langle \delta_{s(x)\al}\rangle_q=\frac{1}{q}\,,\nonumber\\
&& \langle \delta_{s(x)\al}\delta_{s(y)\al}\rangle_q=\frac{1}{q}\,g_q(x,y)+
\frac{1}{q^2}\,[1-g_q(x,y)]\,,
\eea
where $g_q(x,y)$ is the probability that $x$ and $y$ belong to the same 
cluster. Using the definition (\ref{sigma}) we immediately find
\EQ
\langle\sigma_\al(x)\sigma_\al(y)\rangle_q=\frac{q-1}{q^2}\,g_q(x,y)\,,
\EN
from which we see that the pair connectivity in percolation can be obtained as
\EQ
g(x,y)=\lim_{q\goto 1}\frac{1}{q-1}\,\langle\sigma_\al(x)\sigma_\al(y)\rangle_q
\,\,.
\label{connectivity}
\EN
Comparison with (\ref{p}), (\ref{size}) and (\ref{xitilde}) provides the 
following relations with the magnetisation, susceptibility and correlation 
length in the Potts model
\bea
&& P=\lim_{q\goto 1}\frac{M}{\sqrt{q-1}}\,,\nonumber\\
&& S=\lim_{q\goto 1}\frac{\chi}{q-1}\,,\nonumber\\
&& \tilde{\xi}=\xi|_{q=1}\,\,.
\eea
Together with (\ref{number}), they imply the following critical behaviour 
near the percolation threshold
\bea
&& \frac{\langle N_c\rangle}{N}\simeq\tilde{A}_\pm|p_c-p|^{2-\al}\,,\nonumber\\
&& P\simeq\tilde{B}(p-p_c)^\beta\,,\nonumber\\
&& S\simeq\tilde{\Gamma}_\pm|p_c-p|^{-\gamma}\,,\nonumber\\
&& \tilde{\xi}\simeq\tilde{\xi}_0^\pm|p_c-p|^{-\nu}\,,
\eea
where the critical exponents are those of the Potts model evaluated at $q=1$,
and the amplitudes are related to the Potts amplitudes as\footnote{Notice
that the labels $+$ and $-$ refer to $p<p_c$ and $p>p_c$, respectively.}.
\bea
&& \tilde{A}_\pm=\lim_{q\goto 1}\frac{A_\pm\tau_0^{2-\al}}{(q-1)\al(1-\al)
(2-\al)}\,,\nonumber\\
&& \tilde{B}=\lim_{q\goto 1}\frac{B\tau_0^\beta}{\sqrt{q-1}}\,,\nonumber\\
&& \tilde{\Gamma}_\pm=\lim_{q\goto 1}\frac{\Gamma_\pm\tau_0^{-\gamma}}{q-1}\,,
\nonumber\\
&& \tilde{\xi}_0^\pm=\left.\xi_0^\pm\tau_0^{-\nu}\right|_{q=1}\,;
\eea
here we introduced the non-universal positive constant $\tau_0$ entering the 
relation $\tau\simeq\tau_0(p_c-p)$. It follows from these relations that the 
following combinations of critical amplitudes in percolation
\bea
&& \tilde{A}_+/\tilde{A}_-=\lim_{q\goto 1}A_+/A_-\,,\nonumber\\
&& \tilde{\Gamma}_+/\tilde{\Gamma}_-=\lim_{q\goto 1}\Gamma_+/\Gamma_-\,,
\nonumber\\
&& \tilde{\xi}_0^+/\tilde{\xi}_0^-=\xi^+_0/\xi_0^-|_{q=1}\,,\nonumber\\
&& \tilde{R}_C\equiv\al(1-\al)(2-\al)\tilde{A}_+\tilde{\Gamma}_+/\tilde{B}^2=
\lim_{q\goto 1}\frac{R_C}{q-1}\,,\nonumber\\
&& \tilde{R}_\xi^+\equiv[\al(1-\al)(2-\al)\tilde{A}_+]^{1/d}\tilde{\xi}_0^+=
\lim_{q\goto 1}\frac{R_\xi^+}{(q-1)^{1/d}}\,,
\eea
are universal and can be computed from the $q\goto 1$ limit of the Potts 
amplitude ratios. 

Let us see which results we can obtain for percolation in $d=2$ from our study
of the $q$-state Potts model of the previous sections. From (\ref{a}) we 
find
\EQ
\tilde{A}_\pm=-\left.\frac{m_0^2\tau_0^{2-\al}}{2\sqrt{3}}\right|_{q=1}\,\,.
\label{atilde}
\EN
The negative sign of this amplitude agrees with the series result of Domb and
Pearce \cite{DP} which is listed in Table\,4 together with other series and 
Monte Carlo estimates of percolation amplitudes\footnote{The existing 
$\varepsilon$-expansion results are unreliable in two-dimensions since the 
upper critical dimension in the problem is $d_c=6$.}. The equality of 
$\tilde{A}_+$ and $\tilde{A}_-$ follows from duality which is also a crucial 
ingredient of the Domb and Pearce lattice calculation\footnote{The value 
$\tilde{A}_+/\tilde{A}_-=-1$ is quoted in Refs.\,\cite{Aharony} and \cite{PHA}.
We do not understand the origin of this discrepancy.}.
When combined with the value of $\xi_0^+$ quoted in Table\,2, (\ref{atilde}) 
gives the result ($\al=-2/3$ at $q=1$)
\EQ
\tilde{R}_\xi^+\simeq 0.926\,,
\EN
which we think substantially improves the value around $1.1$ one can extract 
from the lattice amplitudes in Table\,4.

We cannot compute directly the ratios involving the amplitudes $\tilde{B}$, 
$\tilde{\Gamma}_-$ and $\tilde{\xi}_0^-$ simply because we are not able to 
solve the functional equation (\ref{omega}) for $q=1$. What we can do is to 
attempt a naive quadratic extrapolation at $q=1$ of the results obtained 
for $q=2,3,4$. Since all the results following from the scattering theory are
analytic in $\lambda$, we perform a quadratic extrapolation in this 
variable\footnote{It is immediately checked that extrapolating in $q$ gives 
essentially the same results for $\tilde{\xi}_0^+/\tilde{\xi}_0^-$ and 
$\tilde{\Gamma}_+/\tilde{\Gamma}_-$ while the value of $\tilde{R}_C$ gets 
modified by less than 10\%.}. From the values of Table\,2 one extrapolates 
$\Gamma_+m_0^2/B^2\approx 3.5$ at $q=1$, which in turn leads to 
$\tilde{R}_C=40 \Gamma_+m_0^2/27\sqrt{3}B^2\approx 3.0$. Similarly, the results
of Table\,3 lead to $\tilde{\xi}_0^+/\tilde{\xi}_0^-\approx 3.76$ and 
$\tilde{\Gamma}_+/\tilde{\Gamma}_-\approx 74.2$. The extrapolated values for
$\tilde{\xi}_0^+/\tilde{\xi}_0^-$ and $\tilde{R}_C$ compare quite well 
with the Monte Carlo result $\tilde{\xi}_0^+/\tilde{\xi}_0^-=4.0\pm0.5$ of
Ref.\,\cite{Corsten} and with the estimate $\tilde{R}_C\approx 2.7-2.8$ 
we deduce from Table\,4\footnote{Starting apparently from the same lattice 
amplitudes, Aharony \cite{Aharony} finds $\tilde{R}_C\approx 4.1-4.2$ and this 
result is quoted also in \cite{PHA}.}. 

The status of the lattice estimates of the ratio $\tilde{\Gamma}_+/
\tilde{\Gamma}_-$ is extremely controversial. While there exists a substantial
agreement (within 20--30\%) on the value of $\tilde{\Gamma}_+$, the amplitude
$\tilde{\Gamma}_-$ turns out to be very small and difficult to determine. 
This resulted in a series of estimates for the ratio ranging from 14 to 220
(see \cite{PHA})! When a value around 200 seemed to be accepted, the authors
of Ref.\,\cite{Corsten} obtained $\tilde{\Gamma}_+/\tilde{\Gamma}_-=75 
(\frac{+40}{-26})$ as a result of their Monte Carlo analysis (this is the most
recent result known to us).

Since we expect the accuracy of our results at $q=3,4$ to be comparable with 
that found at $q=2$, the main source of uncertainty for the extrapolated 
results is in the extrapolation itself. Comparison between the extrapolated 
and the Monte Carlo values for $\xi_0^+/\xi_0^-$ shows that our error does 
not exceed 10\% in this case. We find quite reasonable to assume the same 
level of accuracy for our prediction on $\Gamma_+/\Gamma_-$.
We summarise in Table\,5 the situation about the amplitude ratios in 
two-dimensional percolation we considered in this paper.

\resection{Conclusion}
In this paper we used the form factor bootstrap approach to compute several 
universal quantities for the $q$-state Potts model and isotropic percolation 
in two 
dimensions. The results have been obtained truncating the spectral series for
the two-point correlators at the level of the two-kink contribution. We showed
by comparison with exact results that this approximation is sufficient to 
provide remarkably accurate values for physical quantities like central 
charge, scaling dimensions and critical amplitudes. In particular, we gave 
the first theoretical prediction of the universal amplitude ratios in the 
$q$-state Potts model for $q=3,4$.

Our ability to provide precise predictions for percolation is limited by the 
difficulty to solve Eq.\,(\ref{omega}) around $q=1$. We gave accurate results
for some of the amplitude ratios and relied for the others on an extrapolation
based on the values obtained at $q=2,3,4$. Our predictions (including those 
from extrapolation) are found to be in good agreement with the existing lattice
estimates when the latter are able to provide reasonably accurate results.
This is not the case for the ratio of the mean cluster size amplitudes above 
and below the percolation threshold, for which the different lattice 
determinations span more than one order of magnitude. In light of this, our 
extrapolated value for this quantity represents substantial progress.

All the results of this paper have been obtained within the form factor 
framework, using the $S$-matrix as the only input and without relying on data 
coming from other approaches (e.g. TBA). This was possible also because we 
computed universal combinations of amplitudes concerning two renormalisation 
group trajectories related by dual symmetry, and then describable through
the same scattering theory. Other universal ratios can be defined which 
involve also amplitudes computed at the critical temperature but in presence of
an external magnetic field. When trying to determine them in integrable field 
theory, one faces the problem of fixing the relative normalisations of 
amplitudes computed through different integrable scattering theories.
It is remarkable that also this problem can be solved using the results of the
TBA \cite{SGmass,Fateev} and some more recent developments \cite{LZ,FLZZ}. 
It was shown in Ref.\,\cite{ratios} through the basic example of the Ising
model how the universal ratios at nonzero magnetic field can be computed 
along these lines.

\vspace{1cm}
{\em Acknowledgements.} The work of J.C. was supported by the EPSRC
Grant GR/J78327.
G.D. was partially supported by the European Union contract FMRX-CT96-0012.

\appendix
\appsection
Consider the matrix element
\EQ
\langle K_{\al\beta}(\th')|\Phi(0)|K_{\beta\al}(\th)\rangle\,\,.
\EN
It can be related by analytic continuation to a matrix element between the 
vacuum and a two-kink state. There are, however, two different ways to perform
the analytic continuation, corresponding to the fact that the final two-kink 
state can be an ``In'' or ``Out'' asymptotic state. This leads to the two
equations
\EQ
\langle K_{\al\beta}(\th')|\Phi(0)|K_{\beta\al}(\th)\rangle=
\langle 0_\al|\Phi(0)|K_{\al\beta}(\th'+i\pi)K_{\beta\al}(\th)\rangle+
2\pi\delta(\th-\th')\langle 0_\al|\Phi|0_\al\rangle\,,
\EN
\EQ
\langle K_{\al\beta}(\th')|\Phi(0)|K_{\beta\al}(\th)\rangle=
\langle 0_\beta|\Phi(0)|K_{\beta\al}(\th)K_{\al\beta}(\th'-i\pi)\rangle+
2\pi\delta(\th-\th')\langle 0_\beta|\Phi|0_\beta\rangle\,,
\EN
where the delta function terms are disconnected parts which take into account
kink-antikink annihilation. While these equations are kinematically identical 
to those one obtains when dealing with ordinary particles, the important 
difference to be noticed is the appearance of two different vacuum states.
Subtracting the second equation from the first, one gets
\bea
\langle 0_\al|\Phi(0)|K_{\al\beta}(\th'+i\pi)K_{\beta\al}(\th)\rangle &=&
\langle 0_\beta|\Phi(0)|K_{\beta\al}(\th)K_{\al\beta}(\th'-i\pi)\rangle
\nonumber\\
& + & 2\pi\delta(\th-\th')(\langle 0_\beta|\Phi|0_\beta\rangle-
\langle 0_\al|\Phi|0_\al\rangle)\,\,.
\eea
As long as $\th\neq\th'$, this is exactly the crossing relation 
(\ref{ffcrossing1}). For $\th=\th'$, the above equation implies the presence 
of a simple pole in the two-kink form factor with residue
\bea
&& \mbox{Res}_{\th=\th'}
\langle 0_\al|\Phi(0)|K_{\al\beta}(\th'+i\pi)K_{\beta\al}(\th)\rangle=
-\mbox{Res}_{\th=\th'}
\langle 0_\beta|\Phi(0)|K_{\beta\al}(\th)K_{\al\beta}(\th'-i\pi)\rangle
\nonumber\\
&& =i(\langle 0_\al|\Phi|0_\al\rangle-\langle 0_\beta|\Phi|0_\beta\rangle)\,\,.
\eea
The presence of a pole in the two-particle form factors when the rapidity
difference equals $i\pi$ is known to reflect the non-locality of the operator 
with respect to the fields which interpolate the asymptotic particles (see e.g.
\cite{YZ}). We see that in the low temperature formalism this amounts to the 
fact that the operator has different expectation values on different vacua.
Among the operators considered in this paper, this is the case of the 
magnetisation $\sigma_\al(x)$ which is non-local with respect to the disorder
operators which create the kinks.

\newpage
\hs
\vspace{25mm}

{\bf Table Caption}

\vspace{1cm}
\begin{description}
\item [Table 1]. Central charge and scaling dimension of the magnetisation
operator in the $q$-state Potts model. The results obtained through the form 
factor approach in the two-kink approximation are shown below the exact 
values. 
\item [Table 2]. Values of some amplitude combinations in the $q$-state Potts 
model as obtained through the form factor approach in the two-kink 
approximation.
\item [Table 3]. Universal amplitude ratios in the $q$-state Potts model.
The exact result for $A_+/A_-$ follows from duality; the other values are 
computed through the form factor approach in the two-kink approximation.
\item [Table 4]. Series and Monte Carlo estimates of amplitudes for bond
percolation on the square lattice and site percolation on the triangular 
lattice. The results marked by the superscript $a$, $b$ and $c$ are taken 
from Refs.\,\cite{DP}, \cite{Stauffer} and \cite{Corsten}, respectively.
\item [Table 5]. Universal amplitude ratios in two-dimensional percolation.
The dagger signals extrapolated values. The results marked by the superscript 
$a$, $d$ and $c$ are taken from Refs.\, \cite{DP}, \cite{PHA} and 
\cite{Corsten}, respectively. The series/MC estimates for $\tilde{R}_C$ and 
$\tilde{R}_\xi^+$ are obtained using the results of Table\,4.

\end{description}

\newpage
\hs
\vspace{25mm}

{\bf Figure Caption}

\vspace{1cm}
\begin{description}
\item [Figure 1]. Schematic representation of the two-body scattering 
amplitudes in the Potts model scattering theory. The continuous lines 
represent a kink, the dotted lines the bound state $B$.
\item [Figure 2]. The three-particle vertices in the Potts model scattering
theory.
\item [Figure 3]. Schematic representation of a two-kink form factor.
\item [Figure 4]. (a) Two-kink form factor of the energy operator at the 
resonant rapidity difference $\th=2i\kappa$; (b) two-kink form factor of the
disorder operator at the resonant rapidity difference $\th=2i\pi/3$.
\item [Figure 5]. Central charge in the $q$-state Potts model. The continuous
line is the exact formula (\ref{c}). For $q\leq 3$ the dotted line gives the
result of the $c$-theorem sum rule computed through the form factor approach 
in the two-kink approximation. For $q>3$ the bound state $B$ enters the 
physical spectrum. The upper dotted branch represents the sum of the kink-kink
contribution (lower dotted branch) and the single particle bound state 
contribution.

\end{description}


\newpage

\begin{center}

\vspace{3cm}
\begin{tabular}{|c|c|c|c|}\hline
$ q $ & $2$        & $3$        & $4$      \\ \hline
$ c $ & $ 1/2 $ & $ 4/5 $  & $ 1 $  \\
$   $ & $ 1/2 $ & $ 0.792 $  & $ 0.985 $\\  \hline
$ x_\sigma $ & $ 1/8 $ & $ 2/15 $  &  $1/8$ \\
$          $ & $ 1/8 $ & $ 0.128 $  &  $0.117$ \\ \hline
\end{tabular}
\end{center}

\begin{center}
{\bf Table 1}
\end{center}

\begin{center}

\vspace{3cm}
\begin{tabular}{|c|c|c|c|c|}\hline
$ q $ & $1$ & $2$        & $3$        & $4$      \\ \hline
$ \Gamma_+m_0^2/B^2 $ & --   & $ 2 $ & $ 0.973 $  & $ 0.476 $  \\
$ \xi_0^+m_0 $ & $1.001$ & $ 1 $ & $ 0.998 $  &  $0.992$ \\ \hline
\end{tabular}
\end{center}

\begin{center}
{\bf Table 2}
\end{center}

\begin{center}

\vspace{3cm}
\begin{tabular}{|c|c|c|c|}\hline
$ q $ & $2$        & $3$        & $4$      \\ \hline
$ A_+/A_- $ & $1$        & $1$        & $1$      \\ 
$ \Gamma_+/\Gamma_- $ & $ 37.699 $ & $ 13.848 $  & $ 4.013 $  \\
$ R_C $ & $ 0.3183 $ & $ 0.1041 $  & $ 0.0204 $\\
$ \xi_0^+/\xi_0^- $ & $ 3.162 $ & $ 2.657 $  &  $1.935$ \\
$ R_\xi^+ $ & $ 0.3989 $ & $ 0.3262 $  &  $0.2052$ \\ \hline
\end{tabular}
\end{center}

\begin{center}
{\bf Table 3}
\end{center}

\newpage

\begin{center}

\vspace{3cm}
\begin{tabular}{|c|c|c|}\hline
               & SQ bond & TR site  \\ \hline
$ \tilde{A}_+$ & $-4.24^a$ & $-4.37^a$ \\
$ \tilde{B}$ & $0.77^b$ & $0.78^b$ \\
$ \tilde{\Gamma}_+$ & $0.134^b$ & $0.128^b$ \\
$ \tilde{\xi}_0^+$ & -- & $0.313^c$ \\ \hline
\end{tabular}
\end{center}

\begin{center}
{\bf Table 4}
\end{center}

\begin{center}

\vspace{3cm}
\begin{tabular}{|c|c|c|}\hline
                      & This work & Series/MC  \\ \hline
$ \tilde{A}_+/\tilde{A}_- $ & $1$ & $1^a$ \\
$ \tilde{\Gamma}_+/\tilde{\Gamma}_- $ & $74.2^\dagger$ & $14-220^d$ \\
$ \tilde{\xi}_0^+/\tilde{\xi}_0^- $ & $3.76^\dagger$ & $4.0\pm 0.5^c$ \\
$ \tilde{R}_C$ & $3.0^\dagger$ & 2.7--2.8 \\
$ \tilde{R}_\xi^+$ & $0.926$ & 1.1 \\ \hline
\end{tabular}
\end{center}

\begin{center}
{\bf Table 5}
\end{center}

\newpage
\begin{figure}
\centerline{
\psfig{figure=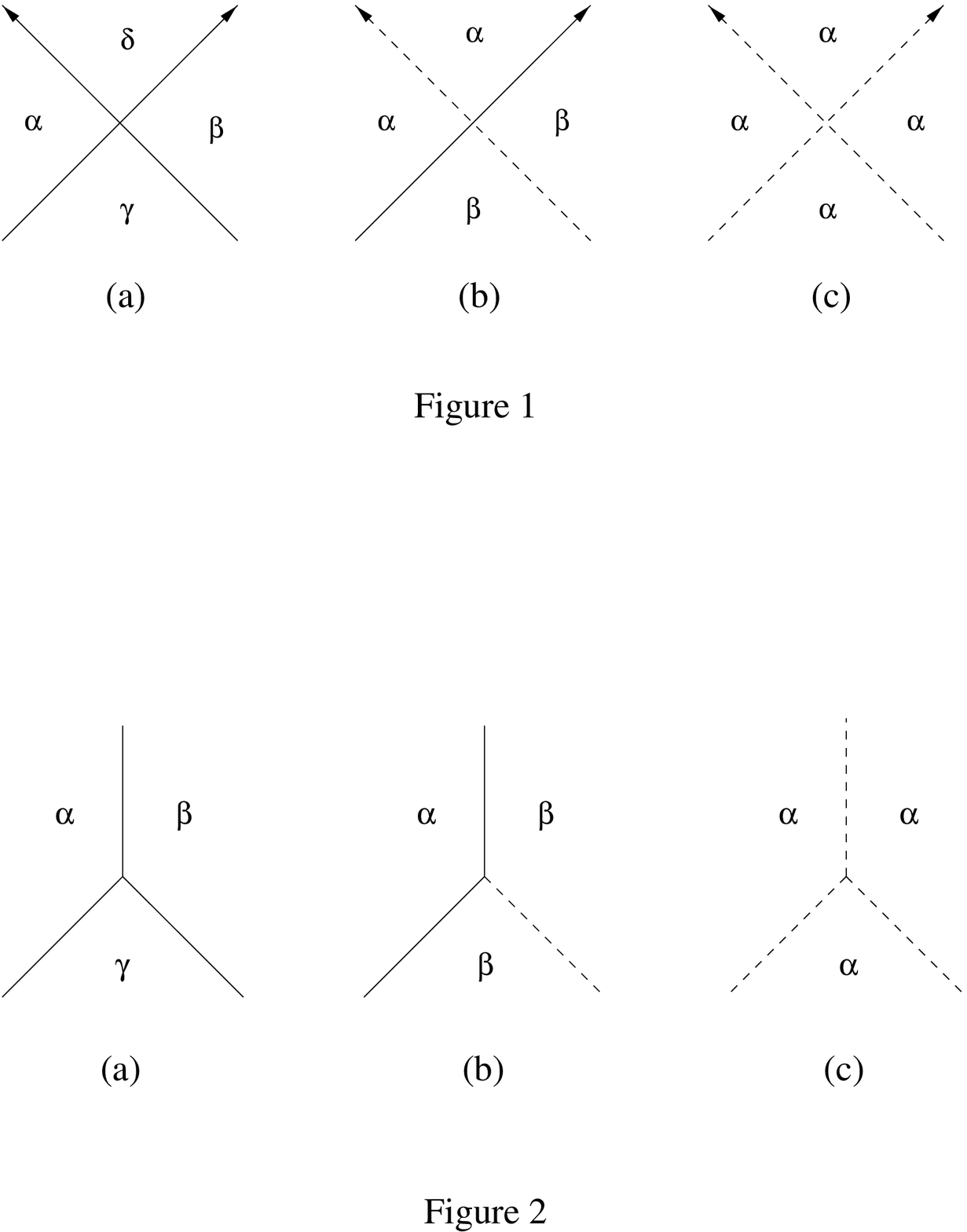}}
\end{figure}

\newpage
\begin{figure}
\centerline{
\psfig{figure=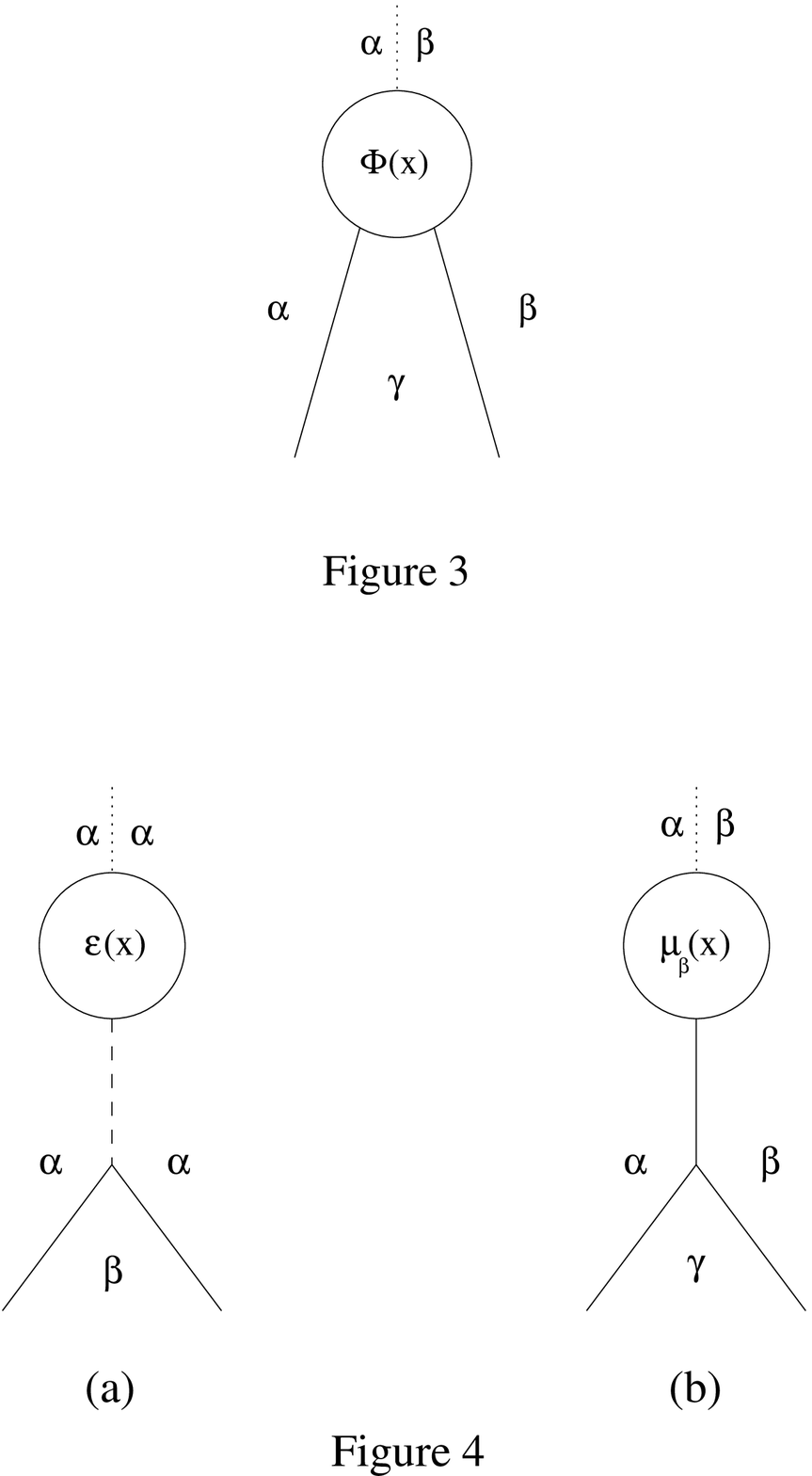}}
\end{figure}

\pagestyle{empty}
\newpage
\begin{figure}
\null\vskip -4cm 
\centerline{
\psfig{figure=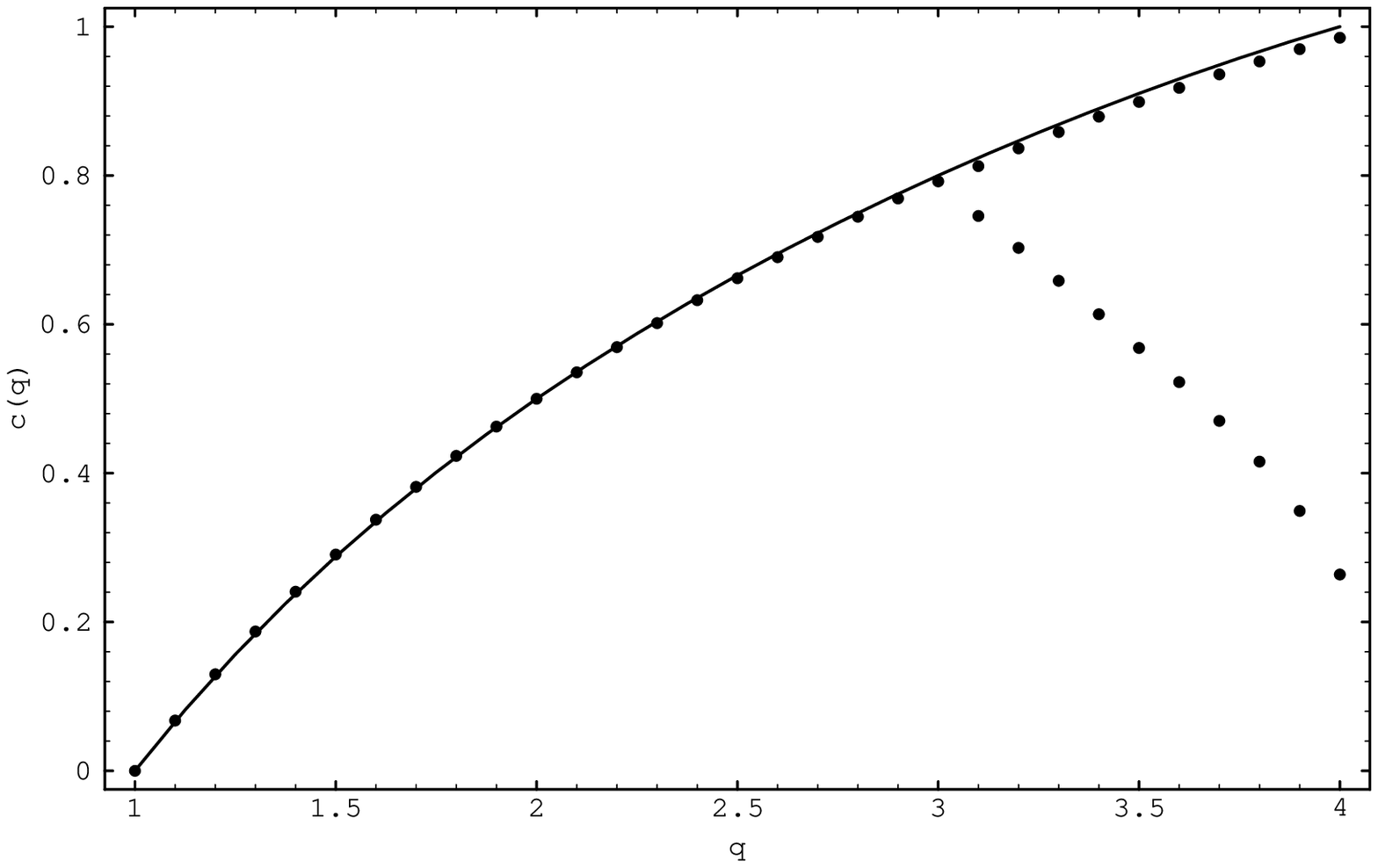}}
\vskip -5cm
\begin{center}
{\bf \large{Figure 5}}
\end{center}
\end{figure}

\end{document}